\documentclass[journal]{IEEEtran}

\usepackage{subfigure}
\usepackage{graphicx}
\usepackage{dcolumn}
\usepackage{bm}
\usepackage{subfigure}
\usepackage{multicol}
\usepackage{graphicx,epsfig,epstopdf}
\usepackage{cite}
\usepackage{color}
\usepackage{amsmath}
\usepackage{subfigure}
\usepackage{graphicx}
\usepackage{dcolumn}
\usepackage{bm}
\usepackage{subfigure}
\usepackage{graphicx,epsfig,epstopdf}
\usepackage{cite}
\usepackage{resizegather}
\usepackage{soul}
\usepackage{kantlipsum,widetext}
\usepackage{cite}
\begin{document}

\title{Reciprocal Metasurfaces for  On-axis Reflective Optical Computing}

\author{\IEEEauthorblockN{Ali Momeni\IEEEauthorrefmark{2},
		Hamid Rajabalipanah\IEEEauthorrefmark{3}, Mahdi Rahmanzadeh\IEEEauthorrefmark{4}, Ali Abdolali\IEEEauthorrefmark{3}\IEEEauthorrefmark{1}, Karim Achouri\IEEEauthorrefmark{5}, Viktar Asadchy\IEEEauthorrefmark{6} and 
	Romain Fleury\IEEEauthorrefmark{2}\IEEEauthorrefmark{1}} \\
\IEEEauthorrefmark{2} Laboratory of Wave
Engineering, School of
Electrical Engineering,
Swiss Federal Institute of
Technology in Lausanne
(EPFL), Lausanne,
Switzerland.\\
\IEEEauthorrefmark{3} Applied Electromagnetic Laboratory, School of Electrical Engineering, Iran University of Science and Technology, Tehran 1684613114, Iran.\\
\IEEEauthorrefmark{4}Electrical Engineering Department, Sharif University of Technology, Tehran 11155-4363, Iran.\\
\IEEEauthorrefmark{5}Nanophotonics and Metrology Laboratory, Swiss Federal Institute of Technology Lausanne (EPFL), 1015
Lausanne, Switzerland.\\
\IEEEauthorrefmark{6}Ginzton Laboratory and Department of Electrical Engineering,
Stanford University, Stanford, California 94305, USA.\\
\IEEEauthorrefmark{1} Corresponding author's email:   abdolali@iust.ac.ir and romain.fleury@epfl.ch
}


\maketitle

\begin{abstract}

Analog computing has emerged as a promising candidate for real-time and parallel continuous data processing. This paper presents a reciprocal way for realizing asymmetric optical transfer functions (OTFs) in the reflection side of the on-axis processing channels. It is rigorously demonstrated that the presence of Cross-polarization Exciting Normal Polarizabilities (CPENP) of a reciprocal metasurface circumvents the famous challenge of Green's function approach in implementation of on-axis reflective optical signal processing while providing dual computing channels under orthogonal polarizations. Following a comprehensive theoretical discussion and as a proof of concept, an all-dielectric optical metasurface is elaborately designed to exhibit the desired surface polarizabilities, thereby  reflecting the first derivative and extracting the edges of images impinging from normal direction. The proposed study offers a flexible design method for on-axis metasurface-based optical signal processing and also, dramatically facilitates the experimental setup required for ultrafast analog computation and image processing.
\end{abstract}


\IEEEpeerreviewmaketitle

\section{Introduction}

\IEEEPARstart{O}{ver} the past few years, the growing demands for ultrafast and large-scale signal, image, and information system processing as well as the saturation of digital computational capacities have brought about
the emergence of new proposals for optical analog computing schemes \cite{goodman2008introduction,stark2012application,mendlovic1993fractional}. In traditional approaches, mathematical processing is mainly performed in the digital domain by using  traditional integrated electronic-based technologies for implementing logics \cite{nakamura2017image,bogoni2005regenerative,brzozowski2001all,mcgeehan2007all}, differentiators and integrators \cite{xu2007all,al1993novel}. 
Unfortunately, these computing hardwares have some major disadvantages related to operational speed and power consumption \cite{jouppi2017datacenter,silva2014performing}. Recently, Silva \textit{et. al} \cite{silva2014performing} have proposed the construction of analog computational metamaterials, based on two distinct approaches: the spatial Fourier transfer and Green’s function (GF) methods \cite{zangeneh2020analogue}.
 \begin{figure*}[t]
	\centering
	\includegraphics[height=3in]{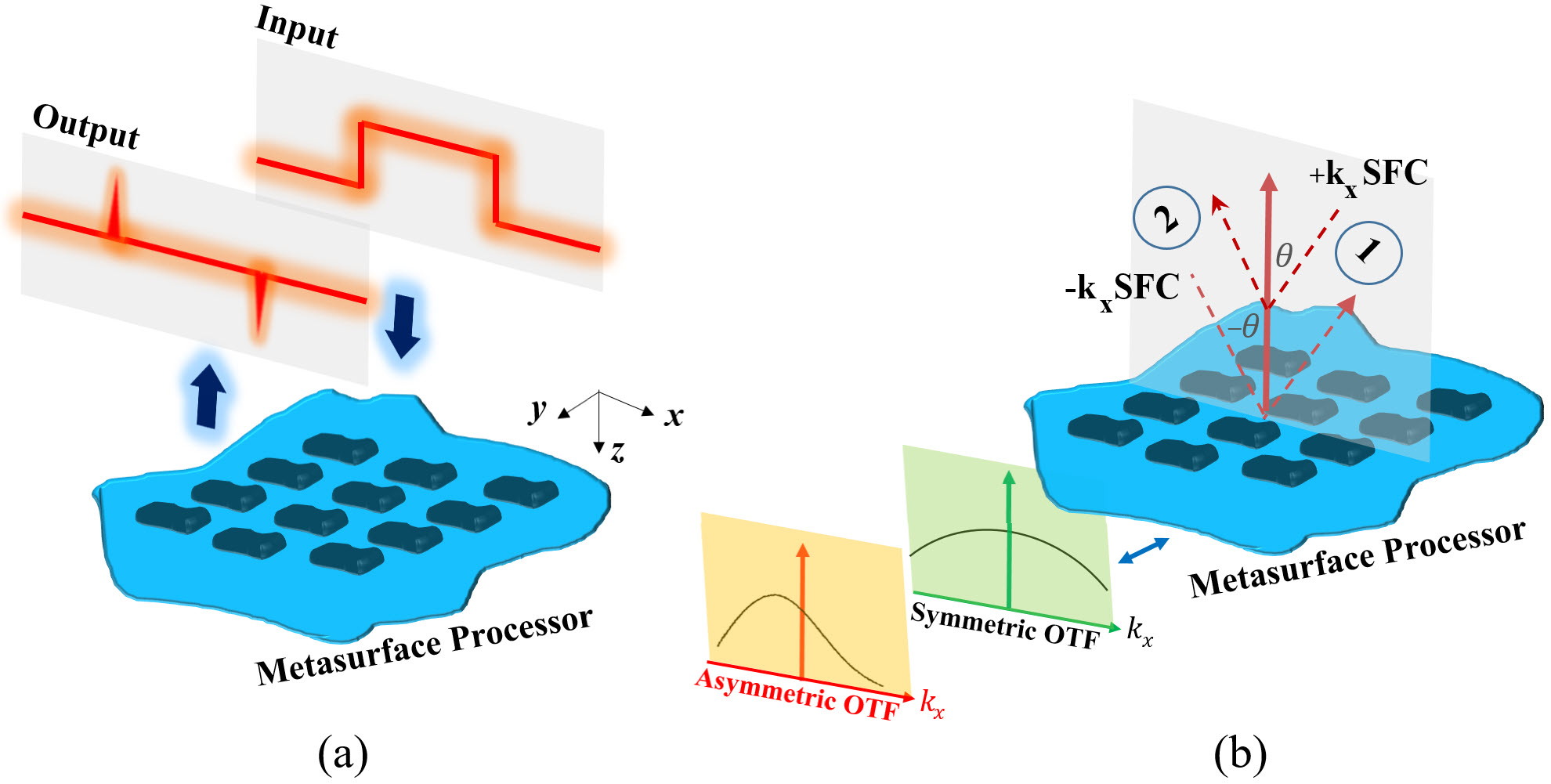}
	\caption{\label{fig:epsart} (a) Schematic sketch of the proposed spatial metasurface processor for performing real-time on-axis reflective analog signal processing (b) Demonstration of the symmetries/asymmetries of the angular scattering around the boresight direction for nonreciprocal/reciprocal metasurfaces.     }
\end{figure*} 
The former solution, executed in the spatial Fourier domain, was initially accompanied with additional bulky optical components (4\textit{f} correlators), hindering miniaturization \cite{silva2014performing}. This shortcoming was later avoided by directly realizing the spatial impulse response of interest by means of the nonlocal transmission or reflection response of artificial structures \cite{zhu2017plasmonic, kwon2018nonlocal, zhou2019optical, kwon2020dual}. A large number of studies presented resonant and non-resonant structures to expand this idea for advanced all-optical signal processing architectures \cite{youssefi2016analog,bykov2018first, zhu2019generalized}. The authors in \cite{kwon2018nonlocal} have demonstrated that the nonlocality of artificial structures can be engineered to enable signal manipulation in the momentum domain over an ultrathin platform for performing basic mathematical operations in transmission mode. \textcolor{black}{Another progress in this area, discussed in \cite{zhu2017plasmonic}, has shown that the interference effects associated with surface plasmon excitations at a single metal-dielectric interface can perform spatial differentiation and edge detection.} More recently, a theoretical work has demonstrated that the Laplacian operator required to do spatial differentiation in transmission mode can be obtained using the guided resonances of a photonic crystal slab  \cite{guo2018photonic}. Besides, different innovative approaches have been also reported, for instance manipulating the complex-valued electromagnetic wave propagating through specially designed recursive paths \cite{estakhri2019inverse}, or the use of topological insulators as a way to increase the robustness to geometrical tolerances \cite{zangeneh2019topological}.

In the GF-based optical signal processing methods, the metasurfaces often aim at realizing transfer functions (TF) with either odd-/or even-symmetric properties (see \textcolor{blue}{Fig. 1a}). For odd/even TFs, the reflection or transmission coefficient of the metasurface processor should be an asymmetric/symmetric function of the incident angle. Due to fundamental limitations arising from the nonlocal behavior of reflection/transmission responses in the wavevector domain, the existing GF-based analog computing proposals have exploited complex oblique illumination setups (challenging to align) to implement odd-symmetric operations \cite{zhou2020analog,youssefi2016analog,zhu2017plasmonic,bykov2018first,abdollahramezani2020meta}. In most practical situations, realizing on-axis signal processing operators with asymmetric optical transfer functions (OTFs) is a key requirement for increasing the compatibility with standard image processing/recognition schemes, such as image sharpening and edge detection \cite{abdolali2019parallel,wan2020optical,kwon2020dual,zhou2020flat}. In addition, reflection-type processing systems are superior to transmission-type ones in terms of compactness. Besides, although using both reflection and transmission channels of the metasurfaces would enhance the degrees of freedom for performing parallel processing, the on-axis reflective channels are still unavailable. The reason is attributed to the fact that realizing an odd-symmetric angular dispersion for co-polarized components of these channels violates the well-known reciprocity theorem  \cite{momeni2018generalized}. Nevertheless, up to now prior proposals for accomplishing on-axis optical signal processing have only considered transmissive configurations \cite{kwon2018nonlocal,momeni2018generalized,abdolali2019parallel,davis2019metasurfaces,zhou2020flat,babaee2020parallel}. 

 \begin{figure*}[t]
	\centering
	\includegraphics[height=1.7in]{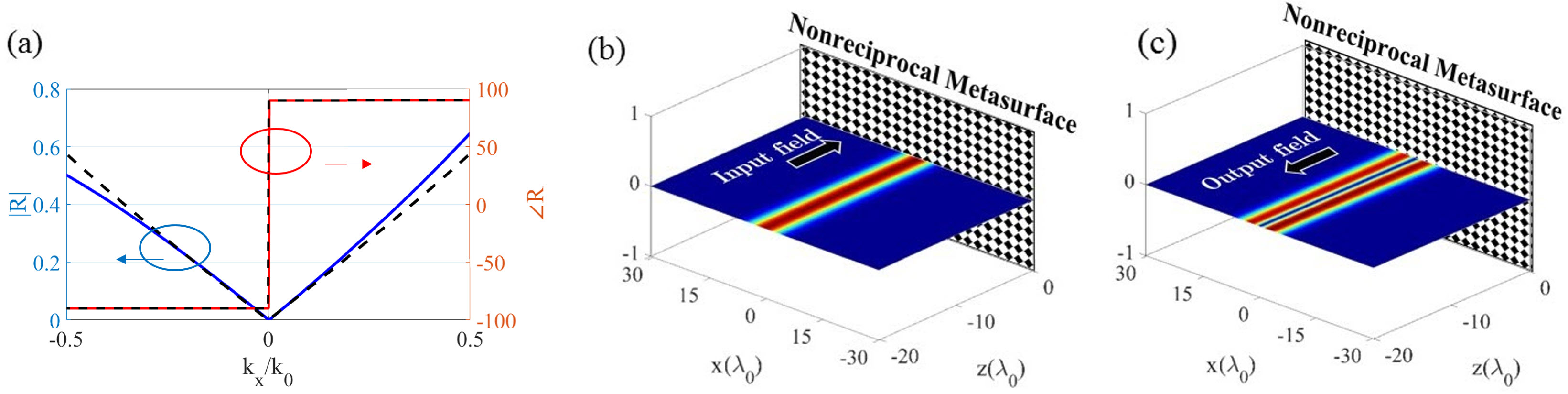}
	\caption{\label{fig:epsart}  (a) The synthesized (dashed line) and exact (solid line) transfer functions obtained by a nonreciprocal anisotropic metasurface differentiator characterized with the following polarizabilities: $\alpha_{\text{ee}}^{xx}=1\times10^{-20}$, $\alpha_{\text{ee}}^{xz}=2.73\times10^{-17}$, $\alpha_{\text{ee}}^{zx}=2.19\times10^{-17}$, $\alpha_{\text{ee}}^{zz}=1.21\times10^{-18}$, and $\alpha_{\text{mm}}^{yy}=1\times10^{-16}$. All quantities are in m$^3$. (b) The Gaussian-shape incident field as well as (c) the derivative reflected field along the boresight direction.     }
\end{figure*}

Motivated by such theoretical developments, here, we propose a reciprocal way for exploiting the on-axis reflective processing channels in bianisotropic metasurfaces. We demonstrate that taking into account the cross-polarization reflections caused by the normal polarizabilities of bianisotropic metasurfaces brings in new degrees of freedom in controlling the meta-atom scattering. We reveal a simple appealing opportunity to circumvent the coercive reflection symmetry in the wavevector domain without resorting to intricate non-reciprocal options or bulky Fourier lenses. We show that using the cross-polarized channels enabled  by a certain group of normal polarizabilities, the odd-symmetric operations can be implemented in on-axis reflective scenarios. Compared with recent works \cite{abdolali2019parallel,momeni2018generalized,babaee2020parallel,momeni2019tunable}, the main contributions of this paper are:\\
  ~~~  1) By considering the general form of the surface polarizabilities of a reciprocal bianisotropic metasurface, we extract the general rules for performing standard on-axis optical signal processing tasks in the reflection side. Some remarks about the transmission configuration are also presented. \\
~~~	2) Realizing an all-dielectric optical metasurface that exposes the desired collective polarzabilites and creates a cross-polarized reflective processing channel.

\section{Theoretical Investigation}

\subsection{ Scattering coefficients of a metasurface}
~~\textcolor{blue}{Fig. 1a} illustrates the employed computational metasurface comprising a uniform array of polarizable meta-atoms in the x-y plane (z=0). Let us consider that \textit{p}-polarized (or \textit{s}-polarized) light with a transverse field-profile of s$_{\text{in}}$(x) impinges normally on the metasurface processor (along the $-$z direction). This incident light is reflected by the metasurface with a beam-profile s$_{\text{out}}$(x) in the $+$z direction. Here, p- and s-polarized refer to the polarized lights whose electric field is oriented along x and y directions, respectively, at normal incidences. In the spatial Fourier domain, each beam profile can be spectrally represented by a superposition of plane waves, as spatial Fourier components (SPF), i.e., s$_{\text{in}}$(x)=$\int\limits_{2W}^{}{\tilde{s}_{\text{inc}}^{\text{}}\left( {{k}_{x}}\right)} ~\text{exp}\,(j{{k}_{z}}{{z}+jk_xx_{\text{}}})\,d{{k}_{x\text{}}},$ and s$_{\text{out}}$(x)=$\int\limits_{2W}^{}{\tilde{H}(k_x)\tilde{s}_{\text{inc}}^{\text{}}\left( {{k}_{x}}\right)} ~\text{exp}\,(-j{{k}_{z}}{{z}+jk_xx_{\text{}}})\,d{{k}_{x\text{}}}$. Here, $W$ denotes the spatial bandwidth of the input signal and k$_x$=k$_0$sin$\theta$ specifies the x component of the wavenumber of the plane wave harmonic illuminating the metasurface with the incidence angle $\theta$. Moreover, $\tilde{H}(k_x)$ refers to the transfer function describing an operator of choice, that we aim at synthesizing. Due to the sub-wavelength periodicity of the metasurface along the x direction, the tangential wave vector k$_x$ must be continuous at the interface. Therefore, the incident plane wave with k$_x$ only generates a reflected propagating plane wave with the same k$_x$. Furthermore, the beam-profile transformation between the incident and reflected lights can be spectrally described by using a spatially-dispersive transfer function representing the mathematical operator of interest, ${\tilde{H}}\left( {{k}_{x}} \right)={{{s}_{out}}\left( {{k}_{x}} \right)}/{{{s}_{in}}\left( {{k}_{x}} \right)}$. Mathematically speaking, the $k_x$-dependency of the transfer function should be imparted by the angular response of the scattering parameter (reflection or transmission coefficient) of the employed metasurface \cite{achouri2019angular}. 
It should be noted that the challenges accompanied with reflective optical signal processing of on-axis illuminations are more severe than those of transmissive configurations. For instance, designing asymmetric reflective OTFs by using co-polarization channels inherently requires non-reciprocity \cite{momeni2018generalized}. Although both reflection and transmission problems will be investigated, the focus  of the paper is the reflection mode. In this case, the output field of the metasurface processor is s$_{\text{out}}$(x)=$F^{-1}$\{s$_{\text{inc}}$(k$_{x}$)$\tilde{R}$(k$_x$)\}. The metasurface can be theoretically treated as a bianisotropic homogeneous sheet in its general form with four distinct sets of collective polarizabilities (${{\overline{\overline{{\hat{\alpha }}}}}_{\text{ee}}}$, ${{\overline{\overline{{\hat{\alpha }}}}}_{\text{em}}}$, ${{\overline{\overline{{\hat{\alpha }}}}}_{\text{me}}}$, ${{\overline{\overline{{\hat{\alpha }}}}}_{\text{mm}}}$) \cite{niemi2013synthesis}.  Keeping the tensorial format of polarizability components in mind, the reflection coefficient (the transfer function) exerted by the metasurface boundary would be different for s- and p-polarized signals, and we write

\begin{equation}
\overline{\overline{{\tilde{H}}}}({{k}_{x}})\equiv \left[ \begin{matrix}
{{{\tilde{R}}}^{s \to s}}\left( {{k}_{x}} \right)\, & {{{\tilde{R}}}^{s \to p}}\left( {{k}_{x}} \right)\,\,  \\
{{{\tilde{R}}}^{p \to s}}\left( {{k}_{x}} \right)\,\, & {{{\tilde{R}}}^{p \to p}}\left( {{k}_{x}} \right)\,  \\
\end{matrix} \right]\,\,\,
\end{equation}
~~where $\tilde{R}$ alludes to the nonlocal reflection coefficient of the bianisotropic metasurface in the wavevector domain. The first and second superscripts refer to the polarization state of the input and output fields, respectively. Indeed, for an arbitrary input signal having a specific polarization state, the bianisotropic metasurface may suggest two possible transfer functions resulting in two different output signals with \textit{s}  or \textit{p} polarization in the reflection mode. Each tensorial component of the transfer function in \textcolor{blue}{Eq. 1} is affiliated to a certain combination of surface polarizabilities, in an explicit or implicit manner. To unveil these dependencies, we need constitutive parameters which do not change with different conditions of the external excitation and only depend on the physical parameters of the metasurface i.e., the shapes and sizes of the meta-atoms. 

With the advent of metasurfaces, a new generation of 2D computational interfaces has emerged to spatially shape the optical fields over deeply sub-wavelength volumes \cite{kuester2003averaged,holloway2012overview,arbabi2017planar}. Several characterization frameworks based on local tensorial scattering coefficients \cite{momeni2018information,rouhi2018real,kiani2020self,kiani2020spatial,rajabalipanah2019asymmetric}, surface impedance formulation \cite{pfeiffer2013metamaterial,selvanayagam2013circuit}, generalized sheet transition conditions (GSTCs) by susceptibility tensors \cite{achouri2015general,achouri2018design} as well as individual/effective polarizabilities \cite{yazdi2017analysis,albooyeh2016electromagnetic,niemi2013synthesis,safari2019electric},  have been fostered to represent metasurfaces in their general bianisotropic forms.
These methods do consider normal components; however, in most studies the normal components are omitted.  However, normal components are pivotal when complex spatially-dispersive behavior are demanded in reflection or transmission mode \cite{achouri2019angular}.  We thus take these components fully into account and follow the so-called Tretyakov-Simovski formalism relating the fields and polarizations in the general form  \cite{albooyeh2016electromagnetic,niemi2013synthesis}
\begin{align}
& \textbf{p}={{\overline{\overline{{\hat{\alpha }}}}}_{\text{ee}}}{{\textbf{E}}_{\text{inc}}}+{{\overline{\overline{{\hat{\alpha }}}}}_{\text{em}}}{{\textbf{H}}_{\text{inc}}} \\ 
& \textbf{m}={{\overline{\overline{{\hat{\alpha }}}}}_{\text{me}}}{{\textbf{E}}_{\text{inc}}}+{{\overline{\overline{{\hat{\alpha }}}}}_{\text{mm}}}{{\textbf{H}}_{\text{inc}}}
\end{align}
~~in which, ${{\overline{\overline{{\hat{\alpha }}}}}_{\text{ee}}}$, ${{\overline{\overline{{\hat{\alpha }}}}}_{\text{em}}}$, ${{\overline{\overline{{\hat{\alpha }}}}}_{\text{me}}}$, and ${{\overline{\overline{{\hat{\alpha }}}}}_{\text{mm}}}$ are called collective surface polarizabilities: electric, magnetoelectric, electromagnetic,
and magnetic ones, respectively. $\textbf{P}$ and $\textbf{M}$ also represent the electric and magnetic polarization densities induced on the metasurface. The transversal components of the reflected and transmitted electric fields can be expressed by \cite{albooyeh2016electromagnetic}: 
\begin{widetext}
\begin{align}
& {{\textbf{E}}_{\text{ref,t}}}=-{{\left( {{\overline{\overline{I}}}_{t}}+{{\overline{\overline{Z}}}_{_{bo}^{to}}}{{\overline{\overline{Y}}}_{_{to}^{bo}}} \right)}^{-1}}\left( {{\overline{\overline{I}}}_{t}}-{{\overline{\overline{Z}}}_{_{bo}^{to}}}{{\overline{\overline{Y}}}_{_{to}^{bo}}} \right)\cdot \,{{\textbf{E}}_{\text{inc,t}}}-{{\left( {{\overline{\overline{I}}}_{t}}+{{\overline{\overline{Z}}}_{_{bo}^{to}}}{{\overline{\overline{Y}}}_{_{to}^{bo}}} \right)}^{-1}}\cdot \left[ j\omega \left( {{\overline{\overline{Z}}}_{_{bo}^{to}}}\cdot {{\textbf{P}}_{t}}\pm\textbf{n}\times {{\textbf{M}}_{t}} \right)\pm j\left( {{\textbf{k}}_{t}}\frac{{{P}_{n}}}{\varepsilon }\mp{{\overline{\overline{Z}}}_{_{bo}^{to}}}\cdot \left( {\textbf{k}_{t}}\times \textbf{n} \right)\frac{{{M}_{n}}}{\mu } \right) \right]\\
& {{\textbf{E}}_{\text{tran,t}}}={{\left( {{\overline{\overline{I}}}_{t}}+{{\overline{\overline{Z}}}_{_{to}^{bo}}}{{\overline{\overline{Y}}}_{_{bo}^{to}}} \right)}^{-1}}\left( {{\overline{\overline{I}}}_{t}}+{{\overline{\overline{Z}}}_{_{to}^{bo}}}{{\overline{\overline{Y}}}_{_{bo}^{to}}} \right)\cdot \,{{\textbf{E}}_{\text{inc,t}}}-{{\left( {{\overline{\overline{I}}}_{t}}+{{\overline{\overline{Z}}}_{_{to}^{bo}}}{{\overline{\overline{Y}}}_{_{bo}^{to}}} \right)}^{-1}}\cdot \left[ j\omega \left( {{\overline{\overline{Z}}}_{_{to}^{bo}}}\cdot {{\textbf{P}}_{t}}\mp \textbf{n}\times {{\textbf{M}}_{t}} \right) \mp j\left( {\textbf{k}_{t}}\frac{{{P}_{n}}}{\varepsilon }\pm{{\overline{\overline{Z}}}_{_{to}^{bo}}}\cdot \left( {\textbf{k}_{t}}\times \textbf{n} \right)\frac{{{M}_{n}}}{\mu } \right) \right]  
\end{align}
\end{widetext}
~~Here, $\textbf{P}$=$\textbf{p}$/$S$ and $\textbf{M}$=$\textbf{m}$/$S$ are polarization surface densities wherein $S$ denotes the unit cell area. Also, the top/bottom sign corresponds to the wave propagating along $\mp$z direction, and $to$ and $bo$ in the subscripts refer to top/bottom medium. For the sake of simplicity, we assume the metasurface is surrounded by free-space. Thus, ${\overline{\overline{Z}}}$ and ${\overline{\overline{Y}}}$=${\overline{\overline{Z}}}^{-1}$ denoting respectively the dyadic impedance and admittance of the corresponding medium, can be written as \cite{tretyakov2003analytical}

\begin{equation}
\overline{\overline{Z}}= \eta\left[ \begin{matrix}
\cos \theta {{\cos }^{2}}\phi +{{{\sin }^{2}}\phi }/{\cos \theta }\,  & \left(\cos \theta -1/\cos \theta \right)\sin \phi \cos \phi \,\,  \\
\left( \cos \theta -1/\cos \theta \right)\sin \phi \cos \phi \,\, & \cos \theta {{\sin }^{2}}\phi +{{{\cos }^{2}}\phi }/{\cos \theta }\,  \\
\end{matrix} \right]\,\,\,
\end{equation}

in which, $\eta$ is the free-space impedance and $\theta$ and $\phi$ are the elevation and azimuth angles of the incident wave. The formulations above may be applied for any special
case of external illumination; for example, s- or p- polarized incident waves or a superposition of these two. To exploit the full potential of the metasurface boundary, each polarizability tensor of \textcolor{blue}{Eqs. (2), (3)} includes both normal and tangential components, i.e., 36 scalar polarizabilities which can be written as 
\begin{equation}
\overline{\overline{{{\alpha }}}}_{\text{ee}}=
\left[ \begin{matrix}
{\alpha _{\text{ee}}^{xx}} & \alpha _{\text{ee}}^{xy}& {\alpha _{\text{ee}}^{xz}}\\ \alpha _{\text{ee}}^{yx}&{\alpha _{\text{ee}}^{yy}}&\alpha _{\text{ee}}^{yz} \\
{\alpha _{\text{ee}}^{zx}}&\alpha _{\text{ee}}^{zy}&{\alpha _{\text{ee}}^{zz}} \\
\end{matrix} \right], 
\overline{\overline{{{\alpha }}}}_{\text{em}}=
\left[ \begin{matrix}
\alpha _{\text{em}}^{xx} & {\alpha _{\text{em}}^{xy}}& \alpha _{\text{em}}^{xz}\\ {\alpha _{\text{em}}^{yx}}&\alpha _{\text{em}}^{yy}&{\alpha _{\text{em}}^{yz}} \\
\alpha _{\text{em}}^{zx}&{\alpha _{\text{em}}^{zy}}&\alpha _{\text{em}}^{zz} \\
\end{matrix} \right]\\ \nonumber \\
\overline{\overline{{{\alpha }}}}_{\text{me}}=
\left[ \begin{matrix}
\alpha _{\text{me}}^{xx} & {\alpha _{\text{me}}^{xy}}& \alpha _{\text{me}}^{xz}\\ {\alpha _{\text{me}}^{yx}}&\alpha _{\text{me}}^{yy}&{\alpha _{\text{me}}^{yz}} \\
\alpha _{\text{me}}^{zx}&{\alpha _{\text{me}}^{zy}}&\alpha _{\text{me}}^{zz} \\
\end{matrix} \right], 
\overline{\overline{{{\alpha }}}}_{\text{mm}}=
\left[ \begin{matrix}
{\alpha _{\text{mm}}^{xx}} & \alpha _{\text{mm}}^{xy}& {\alpha _{\text{mm}}^{xz}}\\ \alpha _{\text{mm}}^{yx}&{\alpha _{\text{mm}}^{yy}}&\alpha _{\text{mm}}^{yz} \\
{\alpha _{\text{mm}}^{zx}}&\alpha _{\text{mm}}^{zy}&{\alpha _{\text{mm}}^{zz}} \\
\end{matrix} \right]
\end{equation}

With substituting \textcolor{blue}{Eqs. 2, 3, 7, 8} into \textcolor{blue}{Eqs. 4, 5} and after some algebraic manipulations, the most general form for the co-polarized reflection and transmission coefficients of the metasurface can be obtained as: 

\begin{widetext}
	\begin{align}
	& R_{\pm}^{s \to s}=\frac{-j\omega }{2}\left[ \frac{\eta }{\cos \theta }\alpha _{\text{ee}}^{yy}\pm \alpha _{\text{em}}^{yx}\mp \tan \theta \alpha _{\text{em}}^{yz}\mp \alpha _{\text{me}}^{xy}-\frac{\cos \theta }{\eta }\alpha _{\text{mm}}^{xx}+\frac{\sin \theta }{\eta }\alpha _{\text{mm}}^{xz}\mp \tan \theta \alpha _{\text{me}}^{zy}-\frac{\sin \theta }{\eta }\alpha _{\text{mm}}^{zx}+\frac{\sin \theta \tan \theta }{\eta }\alpha _{\text{mm}}^{zz} \right] \\ 
	& T_{\pm}^{s \to s}=1-\frac{j\omega }{2}\left[ \frac{\eta }{\cos \theta }\alpha _{\text{ee}}^{yy}\pm \alpha _{\text{em}}^{yx}\mp \tan \theta \alpha _{\text{em}}^{yz}\pm \alpha _{\text{me}}^{xy}+\frac{\cos \theta }{\eta }\alpha _{\text{mm}}^{xx}-\frac{\sin \theta }{\eta }\alpha _{\text{mm}}^{xz}\mp \tan \theta \alpha _{\text{me}}^{zy}-\frac{\sin \theta }{\eta }\alpha _{\text{mm}}^{zx}+\frac{\sin \theta \tan \theta }{\eta }\alpha _{\text{mm}}^{zz}\right] \\
	& R_{\pm}^{p \to p}=\frac{-j\omega }{2}\left[ \eta \cos \theta \alpha _{\text{ee}}^{xx}\mp \alpha _{\text{em}}^{xy}-\eta \sin \theta \alpha _{\text{ee}}^{xz} \pm \alpha _{\text{me}}^{yx}-\frac{1}{\eta \cos \theta }\alpha _{\text{mm}}^{yy}\mp \tan \theta \alpha _{\text{me}}^{yz}+\eta \sin \theta \alpha _{\text{ee}}^{zx}\mp \tan \theta \alpha _{\text{em}}^{zy}-\eta \sin \theta \tan \theta \alpha _{\text{ee}}^{zz}\right] \\ 
	& T_{\pm}^{p \to p}=1-\frac{j\omega }{2}\left[\eta \cos \theta \alpha _{\text{ee}}^{xx}\mp \alpha _{\text{em}}^{xy}-\eta \sin \theta \alpha _{\text{ee}}^{xz} \mp \alpha _{\text{me}}^{yx}+\frac{1}{\eta \cos \theta }\alpha _{\text{mm}}^{yy}\pm \tan \theta \alpha _{\text{me}}^{yz}-\eta \sin \theta \alpha _{\text{ee}}^{zx}\pm \tan \theta \alpha _{\text{em}}^{zy}+\eta \sin \theta \tan \theta \alpha _{\text{ee}}^{zz}\right] 
	\end{align}
\end{widetext}

These equations reveal the angular dispersion behavior of the reflection and transmission coefficients for a non-reciprocal bianisotropic metasurface in its general form with 36 tonsorial components and will be instrumental for the  discussions presented in the following sections.

\subsection{ Nonreciprocal on-axis differentiator}

\begin{figure*}[t]
	\centering
	\includegraphics[height=2.4in
	]{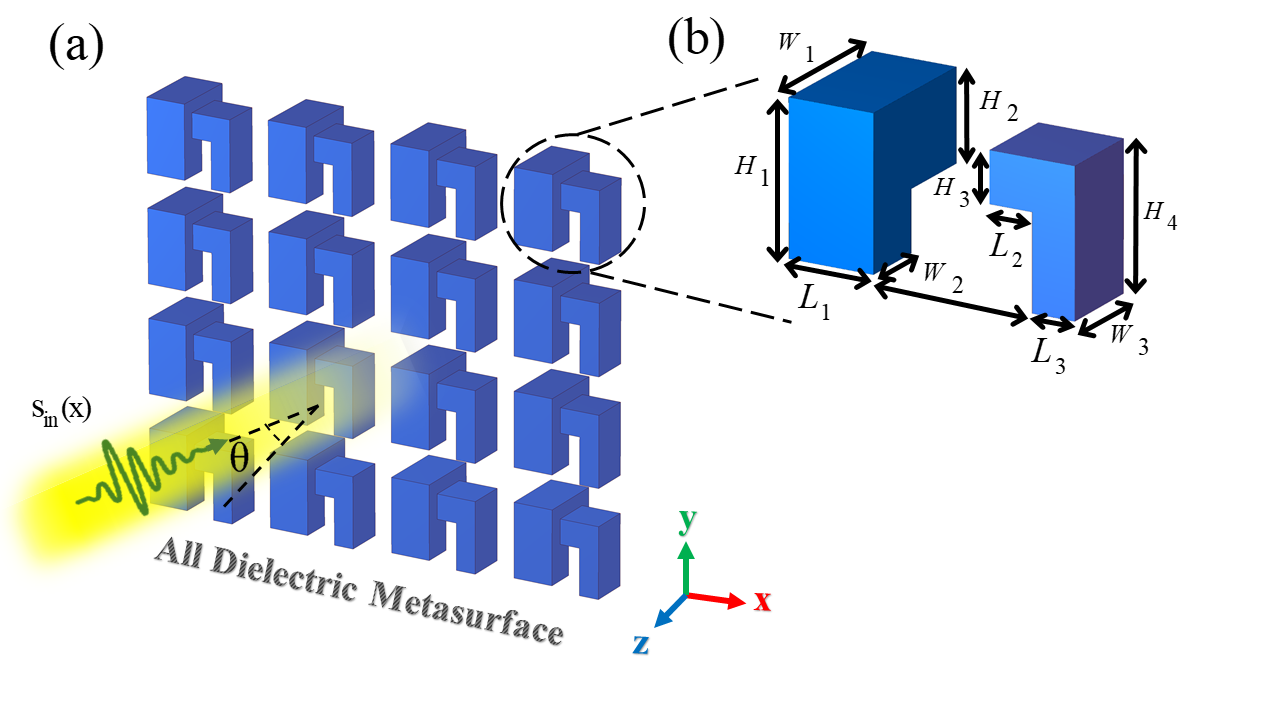}
	\includegraphics[height=2.4in
	]{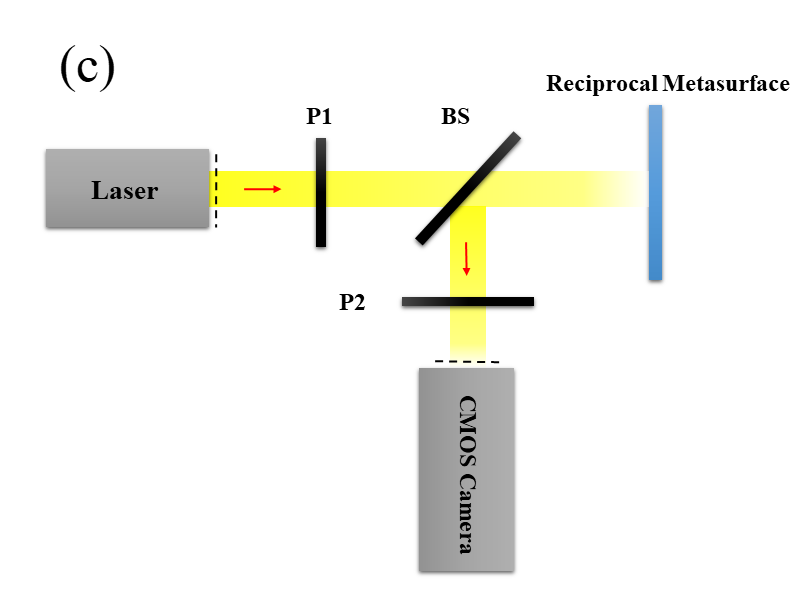}
	\caption{ (a) The array of polarizable meta-atoms for performing optical signal processing. (b) The proposed reciprocal bianisotropic meta-atom exposing normal polarizability components. The geometry was chosen in such a way that both symmetries in xy- and yz- planes are broken. The geometrical parameters are L$_1$=410 nm, L$_2$=205 nm, L$_3$=205 nm, L$_4$=350 nm, w$_1$=695 nm, w$_2$=315 nm, w$_3$=410 nm, H$_1$=760 nm, H$_2$=455 nm, H$_3$=250 nm, and H$_4$=730 nm. (c) The schematic setup for testing the optical signal processing performance of the reflective metasurface in which P and BS stand for the polarizer and beam splitter, respectively. }\label{fig:epsart}
\end{figure*}
The main goal of this paper is to search for the best group of polarizability components whereby the reflection or transmission coefficient in \textcolor{blue}{Eqs. (9)-(12)}, $\tilde{\textbf{R}}$(k$_x$) or $\tilde{\textbf{T}}$(k$_x$), obeys the angular dispersion of the desired mathematical operator. The first-order differentiation operation is a good representative for the family of asymmetric OTFs. Referring to $\tilde{H}$(k$_x$)=jk$_x$ for the first-order differentiation operator, our purpose is to realize an on-axis reflective channel whose transfer function, $\tilde{R}$(k$_x$), has indispensably a non-local odd-symmetric treatment around $\theta=0$ without polarization rotation, i.e., R($\theta$)=R($-\theta$).  Let us consider the conceptual illustration of \textcolor{blue}{Fig. 1b} in which the possibility of R($\theta$)$\ne$R($-\theta$) is graphically investigated. A first-order, polarization-preserving, differentiation operation requires that the co-polarized reflection coefficient imparted on the plane wave coming from $R_{2\to1}$ channel ($\theta$ illumination) essentially differs from that seen from $R_{1\to2}$ channel ($-\theta$ illumination), which is  prohibited by  reciprocity. So, resorting to a non-reciprocal metasurface is inevitable to implement a polarization-preserved first-order differentiator for on-axis illuminations \cite{momeni2018generalized}. This observation can be verified through \textcolor{blue}{Eqs. (9), (11)} in which the necessary condition to keep terms with odd functionality from $\theta$ is $\alpha _{\text{em}}^{yz} \ne -\alpha _{\text{me}}^{zy}$ or $\alpha _{\text{mm}}^{xz} \ne \alpha _{\text{mm}}^{zx}$ for s-polarized incidences and $\alpha _{\text{ee}}^{xz} \ne \alpha _{\text{ee}}^{zx}$ or $\alpha _{\text{me}}^{yz} \ne -\alpha _{\text{em}}^{zy}$ for p-polarized incidences. These constraints for  polarization-preserved reflective optical processing obviously break the Casimir-Onsager reciprocity relations \cite{serdiukov2001electromagnetics}. However, as can be directly inferred from \textcolor{blue}{Eqs. (10), (12)},  polarization-preserving on-axis asymmetric OTFs in transmission regime can be simply implemented by using a reciprocal metasurface having non-zero $\alpha _{\text{mm}}^{xz}$ and $\alpha _{\text{ee}}^{xz}$ for s- and p-polarized incidences, respectively. 

As a representative example, let us consider the following synthesis problem: find the surface polarizabilities of a nonreciprocal metasurface for which a p-polarized input field with an illumination angle of $\theta_{\text{inc}}=0^\circ$ reflects into a co-polarized beam whose profile is the first-order derivative of the one of the input.  A Gaussian-shape beam profile with a spatial bandwidth of $W=0.2k_0$ is considered as the input field. In order to obtain the required surface polarizabilties, the metasurface synthesis will be treated as a non-linear optimization problem which is solved numerically. Firstly, according to the scattering role of each polarizability component, the neutral components are assumed to be zero. The involved polarizabilities are $\alpha_\text{{ee}}^{xx}$, $\alpha_\text{{ee}}^{xz}$, and $\alpha_\text{{ee}}^{zx}$, $\alpha_\text{{ee}}^{zz}$, and $\alpha_\text{{mm}}^{yy}$. Then, we seek for those surface polarizabilities for which the difference between the desired GF, $\tilde{G}_{\text{des}}(k_x)$=jk$_x$, and the angle-dependent reflection coefficient of the metasurface, $\tilde{R}(k_x)$, is minimized within a pre-determined angular range. The cost function is defined as the sum of squares of the differences at a finite number of N samples for both real and imaginary parts $ f={{w}_{\text{re}}}{{\sum\limits_{i=1}^{N}{\left( \operatorname{Re}\left[ {{{\tilde{G}}}_{\text{des}}}\left( {{k}_{x,i}} \right)-{\tilde{R}}\left( {{k}_{x,i}} \right) \right] \right)}}^{2}}+{{w}_{\text{im}}}{{\sum\limits_{i=1}^{N}{\left( \operatorname{Im}\left[ {{{\tilde{G}}}_{\text{des}}}\left( {{k}_{x,i}} \right)-{\tilde{R}}\left( {{k}_{x,i}} \right) \right] \right)}}^{2}}$. Here, $w_{\text{re}}$ and $w_{\text{im}}$ represent the weight coefficients established to selectively adjust the contribution of real and imaginary parts, respectively. At this stage, the metasurface processor is synthesized with the optimized polarizabilities shown in the caption of \textcolor{blue}{Fig. 2}.  It should be noted that, since $\alpha^{\text{xz}}_{\text{ee}}$ is not equal to $\alpha^{\text{zx}}_{\text{ee}}$, the synthesized metasurface is non-reciprocal. The resultant transfer functions along with the input and output fields are displayed in \textcolor{blue}{Figs. 2a-c}. These figures show good agreement between the desired and synthesized transfer functions and output beam profiles. Indeed, the reflected field propagating out along the boresight direction is the 1$^{\text{st}}$-order derivative of the Gaussian-shape input field. The accuracy of differentiation is 99.5\%, described by the Pearson correlation coefficient between the simulated and exact reflected field amplitudes. The results confirm that at the expense of a complex fabrication, a suitably designed nonreciprocal anisotropic metasurface can be thought of as a reflective optical 1$^{\text{st}}$-order differentiator working for input fields coming from normal direction.    

\subsection{ Reciprocal on-axis differentiator}

As an alternative solution, incorporating cross-polarization channels to the problem can elaborately circumvent the nettlesome restrictions arising from reciprocity. In this case, the polarizability components are engineered so that the polarization state of the incident light coming from Port 1 be rotated when it is captured at Port 2 (see \textcolor{blue}{Fig. 1b}). Hereafter, the cross-polarized reflection of $R^{s\to p}$ or $R^{p\to s}$ serves as the transfer function of our processing channel meaning that the output signal has orthogonal polarization with respect to that of the input wave. It should be noted that such a metasurface, if designed, will exhibit  $R_{1\to 2}^{s\to p}$$\ne$$R_{2\to 1}^{s\to p}$ which no longer quashes the reciprocity condition. Similar deduction can be made for $R_{1\to 2}^{p\to s}$$\ne$$R_{2\to 1}^{p\to s}$. Due to the insensitivity of conventional optical sensors, including the human eyes, to the polarization of light, polarization rotation does not degrade the overall performance of optical differentiator, instead, it dramatically facilitates the realization procedure. Based on \textcolor{blue}{Eqs. (2)-(8)}, the cross-polarized reflection and transmission coefficients of the metasurface read as:

\begin{figure*}[t!]
	\centering
	\includegraphics[height=4 in]{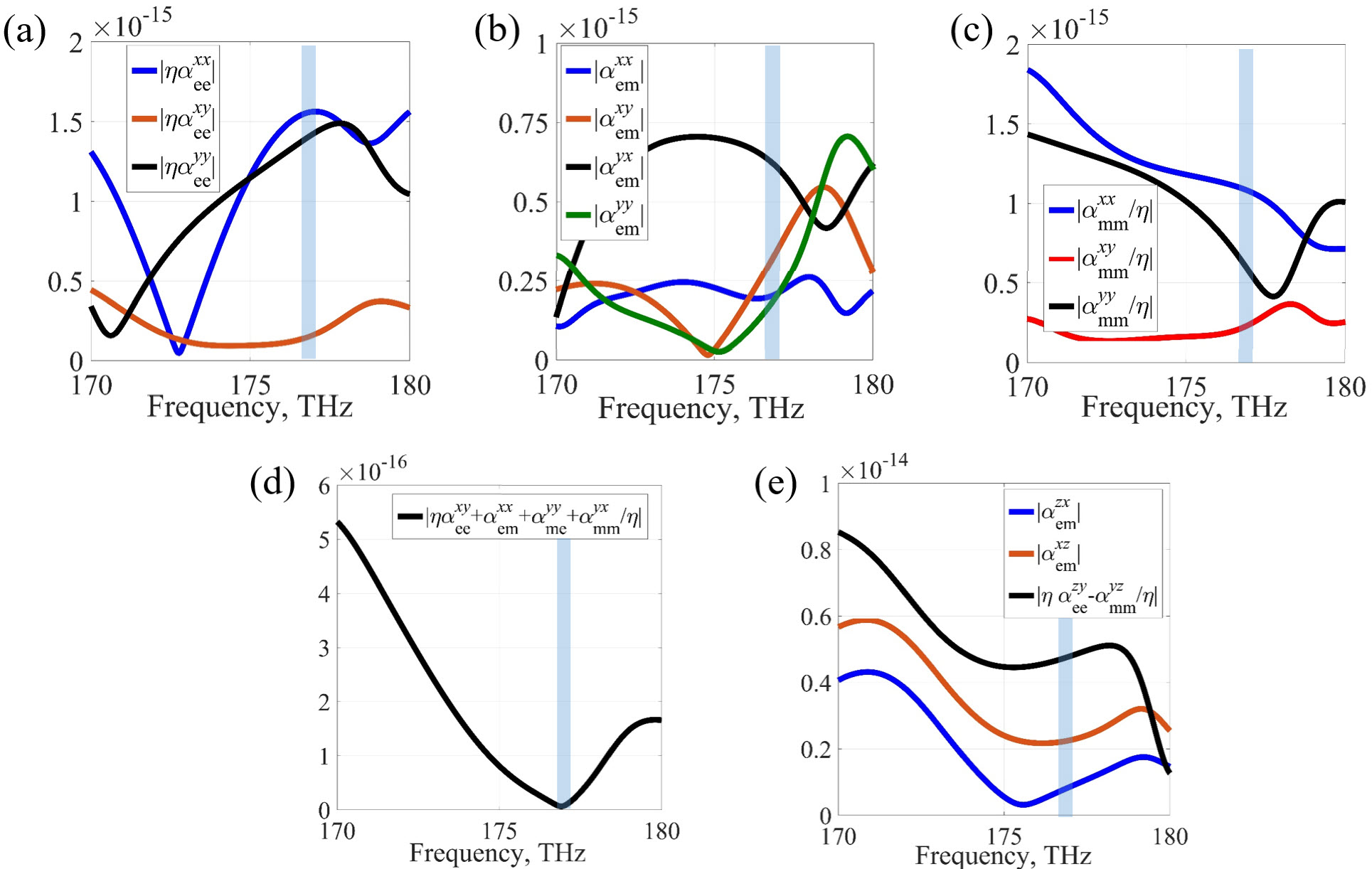}
	\caption{\label{fig:epsart}  (a), (b), (c) The tangential collective polarizabilties of the designed optical metasurface given in \textcolor{blue}{Fig. 3a}. Evaluation of (d) zero cross-polarized reflection and (e) asymmetric angular dispersion conditions on the surface polarizabilities. The working frequency is $f$=177 THz.  }
\end{figure*}

\begin{widetext}
	\begin{align}
	& R_{}^{s \to p}=\frac{-j\omega }{2}\left[\eta \alpha _{\text{ee}}^{xy}\pm \cos \theta \alpha _{\text{em}}^{xx}\mp \sin \theta \alpha _{\text{em}}^{xz}\pm \frac{\alpha _{\text{me}}^{yy}}{\cos \theta }+\frac{1}{\eta }\alpha _{\text{mm}}^{yx}-\frac{\tan \theta }{\eta }\alpha _{\text{mm}}^{yz}+\eta \tan \theta \alpha _{\text{ee}}^{zy}\pm \sin \theta \alpha _{\text{em}}^{zx}\mp \sin \theta \tan \theta \alpha _{\text{em}}^{zz} \right] \\ 
	& T_{}^{s \to p}=-\frac{j\omega }{2}\left[ \eta \alpha _{\text{ee}}^{xy}\pm \cos \theta \alpha _{\text{em}}^{xx}\mp \sin \theta \alpha _{\text{em}}^{xz}\mp \frac{\alpha _{\text{me}}^{yy}}{\cos \theta }-\frac{1}{\eta }\alpha _{\text{mm}}^{yx}+\frac{\tan \theta }{\eta }\alpha _{\text{mm}}^{yz}-\eta \tan \theta \alpha _{\text{ee}}^{zy}\mp \sin \theta \alpha _{\text{em}}^{zx}\pm \sin \theta \tan \theta \alpha _{\text{em}}^{zz} \right] \\
	& R_{}^{p \to s}=\frac{-j\omega }{2}\left[ \eta \alpha _{\text{ee}}^{yx}\mp \frac{1}{\cos \theta }\alpha _{\text{em}}^{yy}-\eta \tan \theta \alpha _{\text{ee}}^{yz}\mp \cos \theta \alpha _{\text{me}}^{xx}+\frac{1}{\eta }\alpha _{\text{mm}}^{xy}\pm \sin \theta \alpha _{\text{me}}^{xz}\mp \sin \theta \alpha _{\text{me}}^{zx}+\frac{\tan \theta }{\eta }\alpha _{\text{mm}}^{zy}\pm \sin \theta \tan \theta \alpha _{\text{me}}^{zz}\right] \\ 
	& T_{}^{p \to s}=-\frac{j\omega }{2}\left[ \eta \alpha _{\text{ee}}^{yx}\mp \frac{1}{\cos \theta }\alpha _{\text{em}}^{yy}-\eta \tan \theta \alpha _{\text{ee}}^{yz} \pm \cos \theta \alpha _{\text{me}}^{xx}-\frac{1}{\eta }\alpha _{\text{mm}}^{xy}\mp \sin \theta \alpha _{\text{me}}^{xz} \mp \sin \theta \alpha _{\text{me}}^{zx}+\frac{\tan \theta }{\eta }\alpha _{\text{mm}}^{zy}\pm \sin \theta \tan \theta \alpha _{\text{me}}^{zz} \right] 
	\end{align}
\end{widetext}
 \begin{figure*}[h!]
	\centering
	\includegraphics[height=1.8 in]{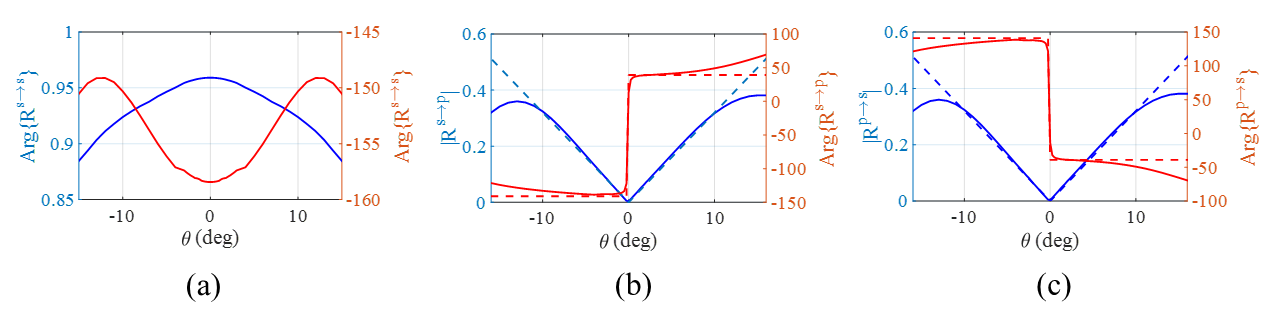}
	\caption{\label{fig:epsart}  The phase and amplitude of (a) co-polarized and (b), s$\to$p and (c) p$\to$s reflection coefficients. The phase and amplitude of the exact transfer function for implementing the first-order differentiation operation are also presented. The solid- and dashed-lines stand for the exact and the synthesized transfer functions, respectively, and the working frequency is $f$=177 THz.  }
\end{figure*}
It should be remembered that we are seeking for those reciprocal collective polarizabity tensors which can be potentially effective to satisfy two important conditions at the same time: exciting the cross-polarization reflected fields and making the cross-polarized reflection to be an asymmetric function of k$_x$ (or $\theta$) variable. As clearly seen from \textcolor{blue}{Eqs. (13), (15)}, although the presence of 9 surface polarizabilities leads to generation of cross-polarized reflection fields for each of s- and p-polarized illuminations, only 4 of them deal with odd-order terms (with respect to $\theta$). Indeed, the necessary condition for reciprocal realization of on-axis asymmetric OTFs in the cross-polarized reflective channel is the presence of \\ \\
rule I:\\

~~$\alpha _{\text{em}}^{xz}$ or $\alpha _{\text{mm}}^{yz}$ or $\alpha _{\text{ee}}^{zy}$ or $\alpha _{\text{em}}^{zx}$ (s-polarized illumination)
\\

~~$\alpha _{\text{ee}}^{yz}$ or $\alpha _{\text{me}}^{xz}$ or $\alpha _{\text{me}}^{zx}$ or $\alpha _{\text{mm}}^{zy}$ (p-polarized illumination) 
\\

This means that the role of normal polarizabilities and bianisotropy is undeniable for a reciprocal metasurface to build asymmetric angular dispersion behavior for the reflection around $k_x$=0. We call these surface polarizability components Cross-polarization Exciting Normal Polarizabilities (CPENP) throughout the paper. The cross-polarized transmission processing channel can also be enabled if the surface polarizabilities $\alpha _{\text{mm}}^{yz}$ or $\alpha _{\text{ee}}^{zy}$ for s-polarized illumination and $\alpha _{\text{ee}}^{yz}$ or $\alpha _{\text{mm}}^{zy}$ for p-polarized input signals contribute to the overall scattering of the metasurface. For the specular case of first-order differentiation operator, $R^{\text{cr}}$$($$k_x$$=0$$)$=0 meaning that no cross-polarized field should exist upon illuminating by on-axis plane waves. Mathematically speaking, \textcolor{blue}{Eqs. (13), (15)} implies that   
\\ \\
rule II:\\

~~$\eta\alpha _{\text{ee}}^{xy}\pm\alpha _{\text{em}}^{xx} \pm\alpha _{\text{me}}^{yy}+\frac{1}{\eta }\alpha _{\text{mm}}^{yx}=0$ (s-polarized illumination)
\\

~~$\eta\alpha _{\text{ee}}^{yx}\mp\alpha _{\text{em}}^{yy} \mp\alpha _{\text{me}}^{xx}+\frac{1}{\eta }\alpha _{\text{mm}}^{xy}=0$ (p-polarized illumination) 
\\

must be satisfied, because otherwise, the metasurface may expose non-zero cross-polarized reflection at $\theta$$=$$0$. In summary, among all possible types of bianisotropy and longitudinal polarizabilities, specific solutions are desired which have been comprehensively discussed above. As an important deduction from \textcolor{blue}{Fig. 1b}, the reciprocity enforces $R^{s\to p}(\theta)$=$R^{p\to s}(-\theta)$ meaning that the mirror version of each mathematical operator realized by $R^{s\to p}$ channel, $\tilde{H}$(k$_x$), will also be constructed by $R^{p\to s}$ channel, $\tilde{H}$($-$k$_x$). Indeed, the proposed approach is inherently dual-polarized and can be applied to both $s$- and $p$-polarized input signals for many types of transfer functions like spatial differentiation and integration for which $\tilde{H}$($-$k$_x$) still yields a usable output wave. This interesting feature which has been rarely reported in the literature may find great potential applications in dual-polarized optical computations \cite{babaee2020parallel,kwon2020dual}.   

\begin{figure*}[h!]
	\centering
	\includegraphics[height=2.9in]{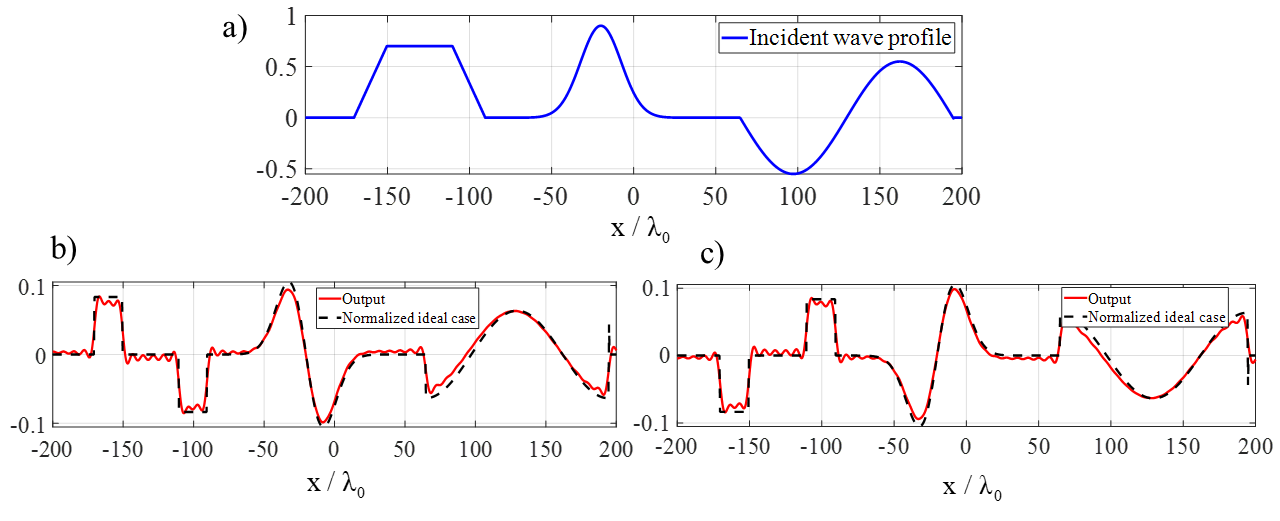}
	\caption{\label{fig:epsart}  (a) Field profile of the input signal, formed by a combination
	of trapezoidal, Gaussian, and sinusoidal functions (b), (c) the s$\to$p and p$\to$s reflected field profiles, respectively. The exact derivative signals have been also presented for the sake of comparison. The metasurface is illuminated from normal direction and the working frequency is $f$=177 THz.  }
\end{figure*} 

\section{Dual-polarized Optical Signal and Image processing}
As known, the macroscopic scattering features of any metasurface is mainly determined by properties of its constituent meta-atoms in the microscopic scale \cite{achouri2018design,albooyeh2016electromagnetic}. In this section, we demonstrate that the proposed all-dielectric optical metasurface shown in \textcolor{blue}{Fig. 3a} exposes the same collective polarizabilities required for performing first-order spatial differentiation at the macroscopic view. The meta-atom of \textcolor{blue}{Fig. 3a} comprises two non-identical L-shaped dielectric particles with a relative 90-degree rotation. The meta-atoms are made of silicon (Si) material whose relative permittivity is extracted from Palik's book \cite{palik1999electronic}. A comprehensive parametric study has been accomplished to search for the best group of parameters making the angular dispersion of the cross-polarized reflection coefficient of the dielectric particles as close to the desired transfer function as possible. The dielectric inclusions are periodically repeated with the periodicity of  1.1 $\mu$m and 0.9 $\mu$m along the both $\text{x}$ and $\text{y}$ directions, respectively. The symmetry/asymmetry properties of the angular scattering at microscopic scale has a straightforward relation with the meta-atom structural symmetries. As a consequence, the geometrical asymmetries of the meta-atoms in both transverse and normal directions yield a reciprocal bianisotropic metasurface whose CPENP are effectively excited at oblique incidences \cite{achouri2019angular,yazdi2017analysis}. In order to implement the first-order differentiator, we design an all-dielectric meta-atom deprived of any geometrical symmetry with respect to $xyz$ coordinates.

Based on the retrieval procedure presented in \cite{yazdi2016polarizability,yazdi2017analysis}, the in-plane collective polarizabilities of the designed metasurface have been extracted. In fact, after the dipole approximation of meta-atoms in a polarizability retrieval setup, the metasurface is illuminated by four normally incident plane waves, and the tangential polarizations can be described in terms of induced dipole moments. By applying an equivalent surface model for the bianisotropic sheet, the \textit{tangential} collective polarizabilities can be calculated as a function of the reflection/transmission coefficients from the array. The retrieved results are displayed in \textcolor{blue}{Figs. 4a-c}. Moreover, the zero cross-polarized reflection condition of the surface polarizabilities at the normal incidence has been assessed in \textcolor{blue}{Fig. 4d}. As seen, the rule II is satisfied in the vicinity of 177 THz at which the metasurface creates no on-axis cross-polarized fields at the reflection side. Hereafter, we intend to distinguish which one of the CPENP is provided by the designed metasurface. After some mathematical manipulations on \textcolor{blue}{Eqs. (13), (15)} and with applying the reciprocity constraints, we have: 

\begin{widetext}
	\begin{align}
& \alpha _{\text{em}}^{xz}=-\alpha _{\text{me}}^{zx}=\frac {1}{j2\omega \sin \theta} \left[\left(R_{\theta,+}^{s\to p}-R_{-\theta,+}^{s\to p} \right)+\left(T_{\theta,+}^{s\to p}-T_{-\theta,+}^{s\to p} \right)\right]
\\
& \alpha _{\text{em}}^{zx}=-\alpha _{\text{me}}^{xz}=\frac {1}{-j2\omega \sin \theta} \left[\left(R_{\theta,+}^{s\to p}-R_{-\theta,+}^{s\to p} \right)+\left(T_{\theta,-}^{s\to p}-T_{-\theta,-}^{s\to p} \right)\right]
\\
& \eta \alpha _{\text{ee}}^{zy}- \frac{1}{\eta} \alpha _{\text{mm}}^{yz}=\frac {1}{-j2\omega \tan \theta} \left[\left(R_{\theta,+}^{s\to p}-R_{\theta,-}^{s\to p} \right)-\left(T_{\theta,+}^{s\to p}-T_{\theta,-}^{s\to p} \right)\right]
\end{align}
	
\end{widetext}

Four s-polarized oblique plane waves illuminating the metasurface along forward and backward directions serve to extract some information regarding the normal components of the collective polarizabilities. By using an arbitrary small incident wave angle $\theta$$=$$4^\circ$, all required scattering parameters in \textcolor{blue}{Eqs. (17)-(19)} have been simulated and the left-hand side parameters are plotted in \textcolor{blue}{Fig. 4e}. As can be seen, the results clearly demonstrate that around 177 THz, among eight possible CPENP, the impact of $\eta \alpha_{\text{ee}}^{zy}-\frac{1}{\eta}\alpha_{\text{mm}}^{yz}$ is more tangible while the role of $\alpha_{\text{em}}^{zx}$ can be neglected. Indeed, the engineered geometry of the designed reciprocal meta-atom makes it as a suitable choice for realizing on-axis asymmetric OTFs in the reflection side. 

For instance, the cross-polarized reflection coefficient of the array can be tailored to mimic the angular trend of R$^{s\to p }(\theta)$=$j\zeta$sin$(\theta)$. The parameter $\zeta$ stands for the gain of differentiator which depends on the value of surface polarizabilites. Through engineering the meta-atom geometry, one can achieve a cross-polarized response that approximates the nonlocal behavior of the first-derivative operation with a phase jump at k$_x$=0. The phase and amplitude of the co- and cross-polarized reflection coefficients at \textit{f}=177 THz are plotted in \textcolor{blue}{Figs. 5a-c}, respectively. The results are numerically recorded by using CST full-wave commercial software, where periodic boundary conditions are applied to the x- and y-directed walls and Floquet ports are considered along the z direction. The first deduction is that the co-polarized reflection coefficient possesses a symmetric response with respect to $k_x$ due to the reciprocity theorem (\textcolor{blue}{Fig. 5a}). The second inference is that, as expected from the presented analysis, the cross-polarized reflection for both s- and p-polarized incidences, possesses an odd-symmetric spatial trend around k$_x$=0. Owing to the reciprocity constraint and $\tilde{R}^{s \to p}(k_x)$=$-\tilde{R}^{s \to p}(-k_x)$, the results of \textcolor{blue}{Figs. 5b, c} differ only in the phase sign. The exact amplitude and phase of the transfer function representing the first-order differentiation operation are fairly compared with those synthesized by the designed metasurface. An excellent agreement between the cross-polarization reflection of the designed metasurface and the exact transfer function has been achieved as long as the normalized spatial bandwidth of the input signals lies {within} $|k_x/k_0|$$<$0.25. This range essentially
provides the maximum spatial resolution of input signals that can be correctly processed by the designed optical metasurface.

To inspect the differentiation performance of the employed metasurafce processor, a complex signal which is a combination
of trapezoidal, Gaussian, and sinusoidal functions (see \textcolor{blue}{Fig. 6a}), has been utilized as the beam profile of the on-axis incidence. The details for the schematic setup are shown in \textcolor{blue}{Fig. 3c}. The polarizers are utilized to filter out the unwanted polarization in each case and the beam splitter is aimed at directing the reflected fields to the camera. The cross-polarized reflections egressing the metasurface from s$\to$p and p$\to$s normal processing channels are captured as output signals and the corresponding results are illustrated \textcolor{blue}{Figs. 6b, c}, respectively. The output signals were in this manner: Each plane wave impinging on the metasurface will be reflected with a certain reflection coefficient corresponding to its incident wave angle, $\tilde{R}(k_x)$. In this case, the angular spectrum of the output fields is given by 
\begin{align}
& \tilde{g}_{\text{ch1}}(k_x)=\tilde{R}^{s \to p}(k_x)\tilde{f}_{\text{ch1}}(k_x) \\
& \tilde{g}_{\text{ch2}}(k_x)=\tilde{R}^{p \to s}(k_x)\tilde{f}_{\text{ch2}}(k_x)
\end{align}
in reflection mode. Here, the subscripts denote the number of processing channel. Thus, the synthesized output fields can be numerically calculated as:
\begin{align}
 E_{\text{ref, ch1}}=\int\limits_{k_{0x}-W}^{k_{0x}+W}{\tilde{R}^{s \to p}(k_x)\tilde{f}_{\text{ch1}}^{\text{}}\left( {{k}_{x}}\right)} ~\text{exp}\,(-j{{k}_{x}}{{x}_{\text{}}}-j{{k}_{z}}{{z}_{\text{}}})\,d{{k}_{x\text{}}} \\
  E_{\text{ref, ch2}}=\int\limits_{k_{0x}-W}^{k_{0x}+W}{\tilde{R}^{p \to s}(k_x)\tilde{f}_{\text{ch2}}^{\text{}}\left( {{k}_{x}}\right)} ~\text{exp}\,(-j{{k}_{x}}{{x}_{\text{}}}-j{{k}_{z}}{{z}_{\text{}}})\,d{{k}_{x\text{}}}
\end{align}

As can be seen, the all-dielectric optical metasurface successfully implements the first-order differentiation of the input signal at both output terminations. Indeed, when the input signal is reflected by the metasurface, the flat regimes are filtered out, the linear segments are converted to flat ones, the sine is converted to the cosine one, and finally, the descending and ascending parts of the Gaussian signal are distinguished. To evaluate the performance of our metasurface differentiator, the numerically-obtained results are compared with the exact first-derivative of the input signals in the same figures. The results are exactly the responses expected from a first-derivative operation with only 2$\%$ error according to ${{e}_{f}}={\left( \left\| s_{\text{out}}^{\text{exact}}(x)-s_{\text{out}}^{\text{simul}}(x) \right\| \right)}/{\left\| s_{\text{out} ,x}^{\text{ideal}}(x) \right\|}$. Therefore, the effective role of CPENP in constructing odd-symmetric nonlocal reflection suggests a new reciprocal way to realize first-order spatial differentiation for normal illuminations. It should be mentioned that the designed metasurface is suitable for optical applications \cite{niemi2013synthesis}.    
\begin{figure}[h]
	\centering
	\includegraphics[height=2in]{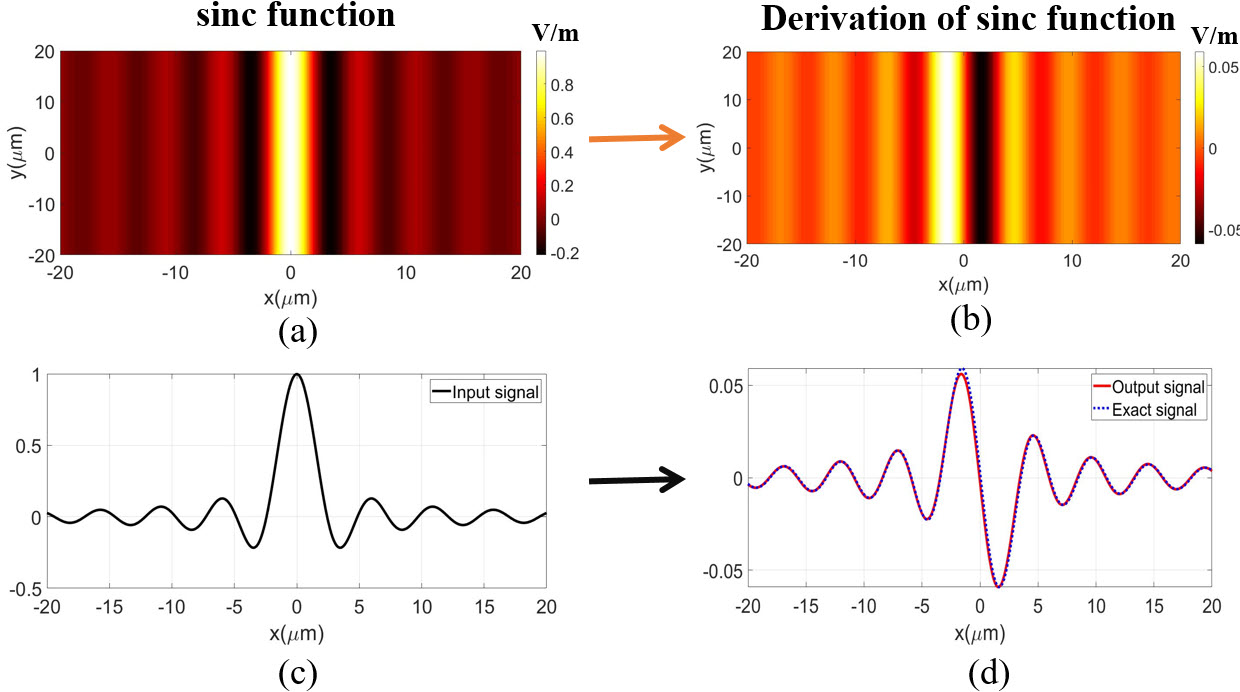}
	\caption{\label{fig:epsart}  (a), (c) The s-polarized sinc-shape incident field profile together with (b), (d) the p-polarized reflected derivative field profile. The exact derivative signals have been also presented for the sake of comparison. The reciprocal metasurface is illuminated from normal direction and the working frequency is $f$=177 THz.  }
\end{figure} 
\begin{figure}[h]
	\centering
	\includegraphics[height=2in]{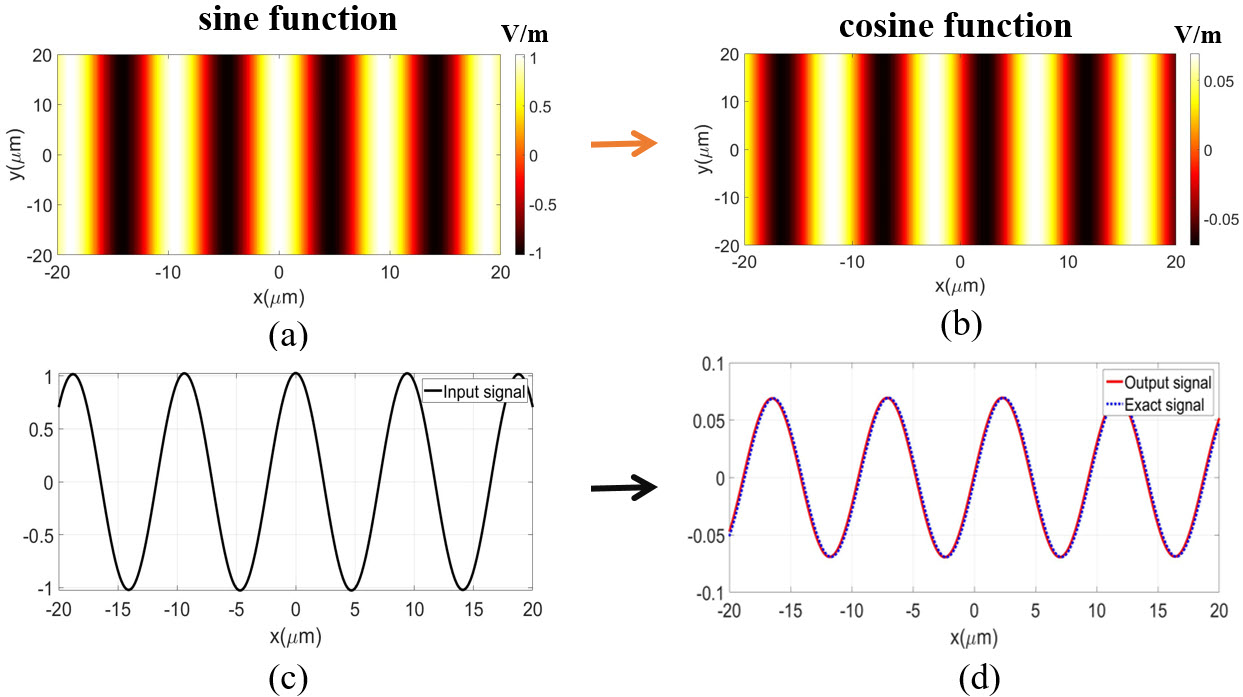}
	\caption{\label{fig:epsart}  (a), (c) The p-polarized sinc-shape incident field profile together with (b), (d) the s-polarized reflected derivative field profile. The exact derivative signals have been also presented for the sake of comparison. The reciprocal metasurface is illuminated from normal direction and the working frequency is $f$=177 THz.  }
\end{figure}

The metasurface processor is dual-polarized and the processing channel can also carry a couple of input/output signals, separately. Two distinct 1D signals (sinc and sinusoidal functions) representing the beam-profiles of s- and p-polarized on-axis incidences are considered as the input signals (see \textcolor{blue}{Figs. 7a, c and Figs. 8a, c}). As can be noticed from \textcolor{blue}{Figs. 7b, d and Figs. 8b, d}, both orthogonal channels can separately serve to construct the first-order derivative of the corresponding input fields at the output termination. 

One of the other advanced applications of the analog optical computing is image processing. Among all possible processing functions, edge detection plays a crucial role in the image segmentation and the other image pre-processing steps \cite{zhou2020flat, abdolali2019parallel, kwon2020dual}. To recognize objects inside a specified image, the spatial differentiation enables us to extract the boundaries between two regions of different intensities \cite{zhou2020flat}. Medical and biological image processing has been gaining significant attention for different medical imaging modalities including X-ray, CT, and MRI \cite{bhattacharyya2011brain,robinson1977edge,zhou2020flat}. Recently, the image processing of biological or ordinary holograms such as various contrast filters and edge detection schemes can be found in literature \cite{zhou2020flat,kwon2018nonlocal,kwon2018nonlocal,momeni2018generalized,abdolali2019parallel}. In this way, we project two differently-polarized images including ordinary and medical ones from normal direction to the designed all-dielectric metasurface differentiator each of which is carried by one of the processing channels (see \textcolor{blue}{Figs. 9a, b}). The reflected images are numerically simulated and the corresponding transverse field profiles are displayed in right side of (see \textcolor{blue}{Figs. 9}). As expected, the 1D edge-detector metasurface successfully exposes all outlines of the normally incident image along the vertical orientations, exhibiting its higher sensibility to fine details as a first-order derivative operator. Besides, it reveals a dual-polarized processing ability. This is while all the prevalent GF-based edge detection schemes fail to operate in on-axis scenarios \cite{zhu2017plasmonic,youssefi2016analog} or need bulky Fourier lenses to do their mission upon illuminating by normally input signals \cite{youssefi2016analog}. \\

\begin{figure}[]
	\centering
	\includegraphics[height=2.1in]{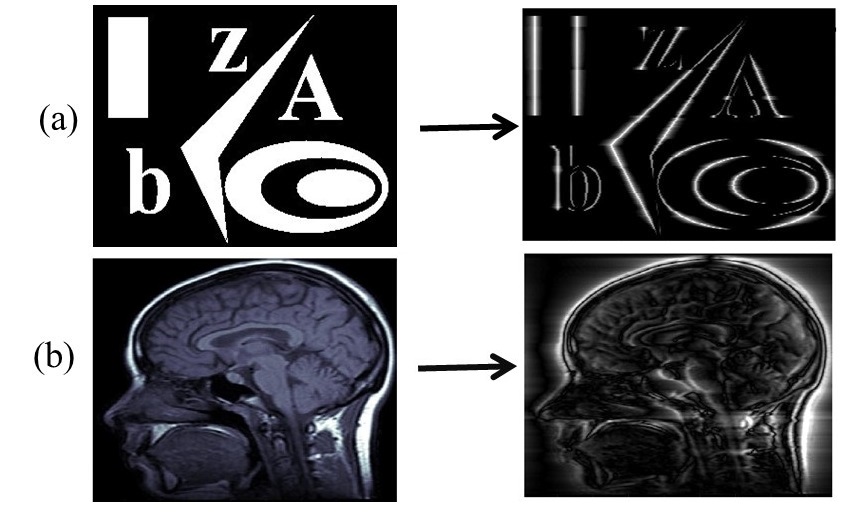}
	\caption{\label{fig:epsart}  Edge Detection of images by exploiting the proposed all-dielectric metasurface processor. (a), (b) Photograph of the s- and p-polarized images whose edges are aimed to be detected. Resulting edge-detected images when 1D all-dielectric metasurface differentiates the input image along x direction, revealing its vertical edges. }
\end{figure}

\section{Conclusion}
~~To conclude, for the first time, we exploited the full macroscopic potential of a reciprocal bianisotropic metasurface to realize dual-polarized asymmetric OTFs at on-axis reflective processing channels. We analytically demonstrated how the cross-polarization channel governed by the CPENP can simply break the angular reflection of the metasurface around $\theta$=0 and mimic the k$_x$-asymmetric spatially-dispersive transfer functions. Motivating by the existing relation between the angular scattering and geometrical mirror symmetries in the microscopic scale, the realization possibility of the proposed design was assessed through proposing an all-dielectric meta-atom and retrieving its important surface polarizabilities. The numerical simulations illustrated that the proposed metasurface reflects the first-derivative of the input signals, with either $s$ or $p$ polarization, coming from the normal direction, without using any bulky Fourier lens. The 1D edge detection performance of the metasurface differentiator was also evaluated where it successfully exposed all boundaries along vertical direction. The analytical design method presented in this paper can be extended to 2D scenarios by involving k$_y$ wavenumber into the relations and obtaining explicit or implicit expression for $\tilde{R}$(k$_x$,k$_y$). Therefore, the designed all-dielectric metasurface offers a simple and ultra-thin normally-oriented channel to perform optical signal/image processes without complicated settings arising from oblique illuminating setups or resorting to nonreciprocal media. Meanwhile, the proposed reflective processing system is more compatible with the current demands on the integrated devices than the transmissive ones while offering a promising vision for realizing parallel computations using both reflection and transmission channels.  \\

Acknowledgements - A.M. and R.F. acknowledge support from the Swiss National Science Foundation (SNSF) under the Eccellenza award number 181232.



\bibliographystyle{IEEEtran}
\bibliography{sample}

\end{document}